\documentclass[12pt]{article}

\usepackage{graphicx}
\usepackage{amssymb}
\usepackage{amsthm}
\usepackage{booktabs}
\usepackage{rotating}
\usepackage{multirow}
\usepackage{eurosym}
\usepackage{amsfonts}
\usepackage{amsmath}
\usepackage{umoline}
\usepackage{caption}
\usepackage{color}
\usepackage{url}
\usepackage{authblk}
\usepackage[english]{babel}

\usepackage{pbox}
\usepackage{footnote}
\usepackage{multirow}
\usepackage{changepage}
\usepackage{tablefootnote}
\usepackage{ccicons}

\usepackage[margin = 2cm]{geometry}

\providecommand{\keywords}[1]{\textbf{\textit{Keywords:}} #1}

\usepackage{hyperref}

\usepackage[natbibapa]{apacite}
\usepackage[symbol]{footmisc}
 
\usepackage{tikz} 
\usepackage{mwe} 
\usepackage{subfigure}
\usepackage{xcolor}
\usepackage[utf8]{inputenc}
\usepackage{import}
\usepackage{float} 
\usepackage{comment}
\usepackage{array}
\newcolumntype{P}[1]{>{\centering\arraybackslash}p{#1}}
\newcolumntype{M}[1]{>{\centering\arraybackslash}m{#1}}
\usepackage{tabularx}

\begin{document}

\title{Assessment of creditworthiness models privacy-preserving training with synthetic data}\footnotetext[1] 
{\scriptsize NOTICE: This is a preprint of a published work. Changes resulting from the publishing process, such as editing, corrections, structural formatting, and other quality control mechanisms may not be reflected in this version of the document. Please cite this work as follows: Muñoz-Cancino, R., Bravo, C., R\'{i}os, S.A., Graña, M. (2022). Assessment of Creditworthiness Models Privacy-Preserving Training with Synthetic Data. In: , et al. Hybrid Artificial Intelligent Systems. HAIS 2022. Lecture Notes in Computer Science(), vol 13469. Springer, Cham. \url{https://doi.org/10.1007/978-3-031-15471-3_32}}
\footnotetext[1]{E-mail addresses: \href{mailto:rimunoz@uchile.cl}{rimunoz@uchile.cl} (Ricardo Muñoz-Cancino), \href{mailto:cbravoro@uwo.ca}{cbravoro@uwo.ca} (Cristi\'{a}n Bravo), \href{mailto:srios@dii.uchile.cl }{srios@dii.uchile.cl } (Sebasti\'{a}n A. R\'{i}os), \href{mailto:manuel.grana@ehu.es}{manuel.grana@ehu.es} (Manuel Graña)}

\author[1]{Ricardo Muñoz-Cancino}
\author[2]{Cristi\'{a}n Bravo}
\author[3]{Sebasti\'{a}n A. R\'{i}os}
\author[4]{Manuel Graña}

\affil[1,3]{Business Intelligence Research Center (CEINE), Industrial Engineering Department, University of Chile, Beauchef 851, Santiago 8370456, Chile}
\affil[2]{Department of Statistical and Actuarial Sciences, The University of Western Ontario,1151 Richmond Street, London, Ontario, N6A 5B7, Canada.}
\affil[4]{Computational Intelligence Group, University of Basque Country, 20018 San Sebasti\'{a}n, Spain.}

\date{}

\maketitle

\begin{abstract}
Credit scoring models are the primary instrument used by financial institutions to manage credit risk. The scarcity of research on behavioral scoring is due to the difficult data access. Financial institutions have to maintain the privacy and security of borrowers’ information refrain them from collaborating in research initiatives. In this work, we present a methodology that allows us to evaluate the performance of models trained with synthetic data when they are applied to real-world data. Our results show that synthetic data quality is increasingly poor when the number of attributes increases. However, creditworthiness assessment models trained with synthetic data show a reduction of 3\% of AUC and 6\% of KS when compared with models trained with real data. These results have a significant impact since they encourage credit risk investigation from synthetic data, making it possible to maintain borrowers’ privacy and to address problems that until now have been hampered by the availability of information.
\end{abstract}


%
%
\section{Introduction}

For decades financial institutions have used mathematical models to determine borrowers’ creditworthiness and consequently manage credit risk. The main objective of these models is to characterize each borrower with the probability of not complying with their contractual obligations \cite{Basel2000}, avoiding to give loans to applicants that will not be able to pay them back. Despite all the years of research on credit scoring, there is still little done on behavioral scoring models, which are the credit scoring models used on those clients who have already been granted a loan, because it requires large volumes of data and a relevant historical depth \cite{Goh2019,Kennedy2013}. In addition, financial institutions are often reluctant to
collaborate in this type of investigation due to concerns about data security and personal privacy. Until now, the use of synthetic data in credit scoring is mainly restricted to balancing the minority class in classification problems using the traditional SMOTE \cite{Gicic2019}, variational autoencoders \cite{Zhiqiang2017}, and lately generative adversarial networks \cite{Fiore2019,Lei2020,Ngwenduna2021}. In these studies, synthetic records of the minority class are generated, and the original data set is augmented. 
In this paper, we present a framework that allows us to train a model on synthetic data and then apply it to real-world data. We also analyze if the model copes with data drift by applying both models to real-world data representing the same problem but obtaining the dataset one year later.
The main findings of our work are:
\begin{itemize}
    \item It is possible to train a model on synthetic data that achieves good performance in real situations.
    \item  As the number of features increases, the synthesized data quality gets worse.
    \item  There is a performance cost for working in a privacy-preserving environment. This cost corresponds to a loss of predictive power of approximately 3\% if measured in AUC and 6\% in KS.
\end{itemize}

\section{Related Work}\label{p03sec:relatedWork}

\subsection{Credit Scoring}

Credit scoring aims to manage credit risk, defined as the potential for a borrower to default on established contractual obligations \cite{Basel2000}. These models intensively use borrower data, demographic information, payment behavior, and even alternative data sources such as social networks \cite{munoz2021,Oskarsdottir2019}, psychometrics \cite{Djeundjie2021}, and geolocation \cite{Simumba2021}.

\subsection{Generative models for synthetic data generation}

Generative models are a subset of machine learning models whose main objective is to learn the real-data distribution and then to generate consistent samples from the learned distribution. Working with synthetic data allows addressing problems where real-data is expensive to obtain, where a large dataset is needed to
train a model, or where the real-data is sensitive or cannot be shared \cite{Torres2018}. For years, statistical methods were the most used ones to estimate the real-world data joint distribution. In this group, Gaussian Mixture Models are the most utilized for this task when there are fewer continuous variables. At the same time, Bayesian Networks are commonly used for discrete variables. The main problem of these methods is dealing with datasets containing numerical and categorical variables. They also present problems when the continuous variables have more than one mode and the categorical variables present small categories \cite{LeiXu2020Thesis}. During the last years, deep learning models have gained popularity to generate synthetic data due to their performance and because they allow us to deal with the problems mentioned above. The generative adversarial networks and the variational autoencoders stand out within these models.

\subsubsection{Generative Adversarial Networks} \label{p03sec:GAN}

Generative adversarial networks are a deep learning framework based on a game theory scenario where a generator network $\mathcal{G}(\cdot)$ must compete with a discriminator network $\mathcal{D}(\cdot)$. The generator network produces samples of synthetic data that attempt to emulate real data. In contrast, the discriminator network aims to differentiate between real examples from the training dataset and synthetic samples obtained from the generator \cite{Goodfellow2016}. Its most basic form, vanilla GAN, $\mathcal{G}(\cdot)$ maps a vector $z$ from a multivariate Gaussian distribution $\mathcal{N}(\mathbf{0},\mathbf{I})$ to a vector $\hat x$ in the data domain $X$. While $\mathcal{D}(\cdot)$ outputs a probability that indicates whether $\hat x$ is a real training samples or a fake sample drawn from $\mathcal{G}(\cdot)$ \cite{LeiXu2020Thesis}. The generator $\mathcal{G}(\cdot)$ and the discriminator $\mathcal{D}(\cdot)$ are alternatively optimized to train a GAN. Vanilla GANs have two main problems, representing unbalanced categorical features and expressing numerical features having multiple modes. To solve this, Xu et al. (2019) \cite{LeiXu2019} present a conditional generator (CTGAN) that samples records from each category according to the log-frequency; this way, the generator can explore all discrete values. Moreover, the multimodal distributions are handled using kernel density estimation to assess the number of modes in each numerical feature.

\subsubsection{Variational autoencoders} \label{p03sec:VAE}

Autoencoders (AE) are an unsupervised machine learning method that enables two main objectives: low-dimensional representation and synthetic data generation. Variational Autoencoder \cite{Kingma2014} interpret the latent space produced by the encoder as a probability distribution modeling the training samples as independent random variables, assuming the posterior distribution defined by the encoder $q_{\theta}(z|x)$ and generative distribution $p_{\phi}(x|z)$ defined by the decoder. To accomplish that the encoder produces two vectors as output, one of means and the other of standard deviations, which are the parameters to be optimized in the model. Xu et al. (2019) \cite{LeiXu2019} present TVAE, a variational autoencoder adaption for tabular data, using the same pre-processing as in CTGAN and the evidence lower bound (ELBO) loss.

\section{Methodology and Experimental Design}\label{p03sec:Methodology}

\subsection{Dataset}\label{p03sec:dataset}
In this work, we use a dataset provided by a financial institution already used for research on credit scoring \cite{munoz2021,munoz2022a}. This dataset includes each borrower financial information and social interactions features over two periods: January 2018 and January 2019; each dataset contains 500,000 individuals. Each borrower is labeled based on their payment behavior in the following 12-month observation period. Each borrower in the 2018 dataset is labeled as a defaulter if it was more than 90 days past due between February 2018 and January 2019 and is labeled as a non-defaulter if it was not more than 90 days past due. Borrowers from the Jan-2019 dataset are similarly tagged. This dataset contains three feature subsets: $X_{Fin}$  corresponds to the borrower’s financial information, $X_{Degree}$ corresponds to the number of connections the borrower has in the social interaction network, and $X_{SocInt}$ are the features extracted from the social interactions.

 \subsection{Synthetic data generation} \label{p03sec:method_syn}
 A step to privacy-preserving credit scoring model building is to generate a synthetic dataset that mimics real-world behavior. In order to accomplish this, we compare the performance of two state-of-the-art synthetic data generators, CTGAN and TVAE, defined in Sect. \ref{p03sec:relatedWork}. The first experiment (\textbf{S01}) only compares these methods using borrowers’ features $X_{Fin}$. The objective of this stage is to find a method to generate synthetic data from real data, and it is not part of this study to find the best way to generate them. Despite not generating an exhaustive search for the best hyper-parameters, we will test two different architectures (Arch) for each synthesizer. Arch A is the default configuration for both methods. In the case of CTGAN, Arch B set up the generator with two linear residual layers and the discriminator with two linear layers, both of size (64, 64). In the case of TVAE Arch B, set hidden layers of (64, 64) for both the encoder and the decoder. Then, in experiment \textbf{S02}, we train a new synthesizer using the best architecture from \textbf{S01}. This experiment uses the borrowers’ features $X_{Fin}$ and exclusively one feature from the network data, the node degree $X_{Degree}$. We only include node degree because its feature enables us to reconstruct an entire network using the random graphs generators. Finally, in experiment \textbf{S03}, the borrowers and social interaction features ($X_{Fin} + X_{Degree} + X_{SocInt}$) are used to train a synthesizer. This experiment corresponds to the traditional approach to generating synthetic data from a dataset using social interaction features.

\subsection{Borrower's creditworthiness assessment} \label{p03sec:method_mod}
The objective of this stage is to have a framework that allows us to estimate the borrower’s creditworthiness from a feature set. This modeling framework is based on previous investigations \cite{munoz2021,munoz2022a}. This stage begins by discarding attributes with low or null predictive power and selecting uncorrelated attributes. The correlation-based selection method begins by selecting the attribute with the highest predictive power. It then discards the possible selections if the correlation exceeds a threshold $\rho$. This step is repeated until no attributes are left to select. To ensure the model generalization capability, we work under a K-fold cross-validation scheme; in this way, the feature selection and the model training use K-1 folds, and the evaluation is carried out with the remaining fold. Additionally, we use two holdout datasets, one generated with information from the same year as the training dataset but not contained. The second contains information from one year later. Both the results of the validation fold and the holdout dataset are stored to use a t-test later to compare different models \cite[Ch.\ 12]{Flach2012}.

\subsection{Evaluation Metrics}\label{p03sec:evaluationmetrics}

In this section, we describe a set of metrics that will help us to evaluate the performance of the synthetic data generators and the classification models used for creditworthiness assessment. The area under the curve \textbf{(AUC)} is a performance measure used to evaluate classification models \cite{Bradley1997}. The AUC is an overall measure of performance that can be interpreted as the average of the true positive rate for all possible values of the false positive rate. A higher AUC indicates a higher overall performance of the classification model \cite{Park2004}. Another classification performance measure is the \textbf{F-measure}. This metric is calculated as the harmonic mean between precision and recall. It is beneficial for dichotomous outputs and when there is no preference between maximizing the model’s precision or recall \cite{Hripcsak2005}. Kolmogorov-Smirnov (KS) statistic measures the distance separating two cumulative distributions \cite{Hodges1958}. The KS statistic ranges between 0 and 1 and is defined as $D = \max_{x}|F_{1}(x) - F_{2}(x)|$, where $F_{1}$ and $F_{2}$ are two cumulative distributions. In the case of creditworthiness assessment, we are interested in the difference between the cumulative distributions of defaulters and non-defaulters, and a higher $D$ indicates a higher discriminatory power. However, in the case of synthetic data generation, we are interested in the real data distribution and the synthetic data distribution being as similar as possible; in this way, a lower $D$ indicates a better synthetic data generation. In order for all the acceptance criteria to be the same, we define the \textbf{KSTest} as $1-D$; in this way, a higher KSTest indicates a better synthetic data generator. In the synthetic data generation problem, the KS is only valid to measure the performance for continuous
features; to handle categorical features, we will use the chi-square test (CS). The CS is a famous test to assess the independence of two events \cite{McHugh2013}. We will call \textbf{CSTest} to the resulting p-value for this test. Therefore a small value indicates we can reject the null hypothesis that synthetic data has the real data distribution. In the synthetic data generation problem, we want to maximize the CSTest.

\subsection{Experimental setup}
The parameters of the univariate selection are set at $ KS_{min} = 0.01 $ and $ AUC_{min} = 0.53 $, i.e., we discard feature with a univarite performance lower than $ KS_{min}$ or $ AUC_{min}$.
In the multivariate selection process, we set $ \rho = 0.7 $ in the process to avoid high correlated features \cite{Akoglu2018}. The N-Fold Cross-Validation stage is carried out considering $ N = 10 $, and in each iteration, the results of regularized logistic regression and gradient boosting \cite{Friedman2001} models are displayed.

\section{Results and Discussion}\label{p03sec:Results}

In this section, we present the results of our methodology. We start with the implementation details. Then, we compare the synthesizers, and finally, we analyze the creditworthiness assessment performance of the models trained using synthetic data.
\subsection{Implementation Details}\label{p03sec:ImpDetails}

In this work, we used the Python implementations of Networkx v2.6.3 \cite{Hagberg2008} and Synthetic Data Vault (SDV) v5.0.0 ~\cite{SDV2016} for networks statistics and synthetic data generation, respectively. To conduct the experiments, we used a laptop with 8 CPU cores Intel i7 and 32 GB of RAM.

\subsection{Synthetic Data Generation Performance}
The first objective is to analyze the performance of the methods to generate synthetic data presented above, CTGAN and TVAE. Table \ref{p03table:Syn_perf} shows the results obtained. The features Synthesizer training features corresponds to the training feature set, while Arq indicates the network architecture defined in Sect. \ref{p03sec:method_syn}.  The experiment S01 consisted in comparing both synthesizer using two different architectures. It is observed that a reduction in the number of layers reduces the execution times considerably in both cases, being TVAE, the one that presented the fastest execution times. KSTest show us the performance to synthesize continuous features, where TVAE achieves better performance than CTGAN. The difference between TVAE architectures is almost negligible when evaluate continuous features performance. The performance to synthesize categorical features is measured using CSTest. In this case, TVAE obtained higher performance again, the differences between architectures is slightly higher to architecture A. Another popular approach to measuring the synthesizer performance is training a classifier to distinguish between real and synthetic data. The column Logistic Detection in Table 1 shows the result after training a logistic regression model; the value displayed corresponds to the complementary F-measure. In this way, values closer to 1 indicate that the classifier cannot distinguish between real and synthetic data, and values closer to 0 mean the classifier efficiently detects synthetic data. It can be seen that TVAE achieve the best performance, but this performance decreases as we include more features to the synthesizer.

\begin{table}[ht]
\footnotesize
\centering
\caption{Synthetic data generators performance}
\label{p03table:Syn_perf}
\resizebox{0.95\textwidth}{!}{%
\begin{tabular}{|c|c|c|c|c|c|c|c|}
\hline
Experiment&Synthesizer training features&Synthesizer&Arch&Exec Time (m)&CSTest&KSTest&Logistic Detection\\ \hline
\multirow{4}{*}{\parbox{1.5cm}{\centering S01}} &\multirow{4}{*}{\parbox{2.5cm}{\centering $X_{Fin}$}} &\multirow{2}{*}{\parbox{2.0cm}{\centering CTGAN}}&A&410&0.836&0.864&0.697\\
& & &B&260&0.861&0.846&0.749\\ \cline{3-8}
& &\multirow{2}{*}{\parbox{2.0cm}{\centering TVAE}}&A&230&0.962&0.868&0.803\\
& & &B&130&0.952&0.861&0.756\\  \hline
S02&$X_{Fin} + X_{Degree}$&TVAE&B&140&0.935&0.836&0.644\\  \hline
S03&$X_{Fin} + X_{Degree} + X_{SocInt} $&TVAE&A&400&0.924&0.809&0.539\\
S03&$X_{Fin} + X_{Degree} + X_{SocInt} $&TVAE&B&320&0.907&0.825&0.542\\
S03&$X_{Fin} + X_{Degree} + X_{SocInt} $&TVAE&B&465&0.930&0.819&0.513\\
\hline
\end{tabular}}
\end{table}

\subsection{Creditworthiness assessment performance on real data}

This section establishes a comparison line for the performance of the models trained with synthetic data. In order to establish this comparison, we first trained classifiers using real-world data and tested their performance using the holdout datasets previously defined. Table \ref{p03table:real_perf} shows the results of training models according to the methodology described in \ref{p03sec:method_mod}. The performance is measured using AUC and KS on the two holdout datasets; the 10-folds mean and its standard deviation are shown for each statistic. For each feature set, we trained two classifiers, logistic regression and gradient boosting. The results show that gradient boosting obtains better results compared to logistic regression. More details of this comparison are shown in Table \ref{p03tbl:real_mod_comp}, where we quantify the higher predictive power of gradient boosting.

\begin{table}[ht]
\footnotesize
\centering
\caption{Creditworthiness assessment performance for models trained on real data}
\label{p03table:real_perf}
\resizebox{0.95\textwidth}{!}{%
\begin{tabular}{|c|c||c|c||c|c|}
\hline
\multirow{2}{*}{\parbox{3.0cm}{\centering Classifier training features}}&\multirow{2}{*}{\parbox{1.5cm}{\centering Classifier}}&\multicolumn{2}{c||}{ {\centering Holdout 2018}}&\multicolumn{2}{c|}{ {\centering Holdout 2019}}\\ \cline{3-6}
&&AUC&KS&AUC&KS\\ \hline
$X_{Fin}$&GB&0.88 $\pm$ 0.001&0.59 $\pm$ 0.002&0.82 $\pm$ 0.001&0.50 $\pm$ 0.002\\
$X_{Fin}$&LR&0.87 $\pm$ 0.001&0.58 $\pm$ 0.001&0.82 $\pm$ 0.001&0.50 $\pm$ 0.002\\ \hline
$X_{Fin}+X_{Degree} + X_{SocInt}$&GB&0.88 $\pm$ 0.001&0.59 $\pm$ 0.002&0.82 $\pm$ 0.001&0.50 $\pm$ 0.002\\
$X_{Fin}+X_{Degree}+ X_{SocInt}$&LR&0.87 $\pm$ 0.001&0.58 $\pm$ 0.002&0.83 $\pm$ 0.001&0.50 $\pm$ 0.002\\ \hline
$X_{Degree} + X_{SocInt}$&GB&0.61 $\pm$ 0.002&0.17 $\pm$ 0.002&0.62 $\pm$ 0.001&0.18 $\pm$ 0.002\\
$X_{Degree} + X_{SocInt}$&LR&0.60 $\pm$ 0.001&0.17 $\pm$ 0.002&0.61 $\pm$ 0.001&0.18 $\pm$ 0.002\\
\hline
\end{tabular}}
\end{table}

Based on the results presented above, we will select gradient boosting for the comparisons against the models trained on synthetic data that we will present in the next section.

\begin{table}[ht]
\footnotesize
\centering
\caption{Gradient boosting and logistic regression comparison on real data (holdout 2018)}
\label{p03tbl:real_mod_comp}
\resizebox{0.85\textwidth}{!}{%
\begin{tabular}{|c|c|c|c|c|}
\hline
Classifier training features&AUC diff (\%)&KS diff (\%)&AUC diff p-value&KS diff p-value\\ \hline
$X_{Fin}$&0.70\%&1.65\%&0.000&0.000\\
$X_{Fin}+X_{Degree}+ X_{SocInt}$&0.84\%&1.91\%&0.000&0.000\\
$X_{Degree} + X_{SocInt}$&1.65\%&2.36\%&0.000&0.000\\

\hline
\end{tabular}}
\end{table}

\subsection{Creditworthiness assessment performance on synthetic data}

This section aims to know how the performance of a creditworthiness assessment model (the classifier) behaves when trained on synthetic data and applied to real-world data. Table \ref{p03tbl:fake_perf1} shows the performance indicators on real-world data. Considering all synthesizers are trained with at least the feature set $X_{Fin}$, the results of training the classifier with $X_{Fin}$ are also displayed for all synthesizers. It is observed that regardless of the synthesizer, training the classifier incorporating at least feature set $X_{Fin}$ produces similar performances in 2018 except in \textbf{S02}. However, when we analyze how much the model degrades, the model trained with synthetic $X_{Fin}$ from synthesizer \textbf{S01} is the one that suffers a minor discrimination power loss. It can be explained in part that a better synthesizer manages to capture better the proper relationship between the borrower features and the default.

\begin{table}[ht]
\footnotesize
\centering
\caption{Creditworthiness assessment performance on real data for model trained on synthetic data}
\label{p03tbl:fake_perf1}
\resizebox{0.95\textwidth}{!}{%
\begin{tabular}{|c|c||c|c||c|c|}
\hline
\multirow{2}{*}{\parbox{2.0cm}{\centering Synthesizer Experiment}}&\multirow{2}{*}{\parbox{4.0cm}{\centering Classifier training features}}&\multicolumn{2}{c||}{ {\centering holdout 2018}}&\multicolumn{2}{c|}{ {\centering holdout 2019}}\\ \cline{3-6}
&&AUC&KS&AUC&KS\\ \hline

S01&$X_{Fin}$&0.85 $\pm$ 0.003&0.53 $\pm$ 0.002&0.82 $\pm$ 0.002&0.48 $\pm$ 0.002\\
S02&$X_{Fin}$&0.82 $\pm$ 0.001&0.51 $\pm$ 0.001&0.80 $\pm$ 0.001&0.46 $\pm$ 0.002\\
S03&$X_{Fin}$&0.85 $\pm$ 0.002&0.55 $\pm$ 0.002&0.80 $\pm$ 0.002&0.46 $\pm$ 0.002\\ \hline

S03&$X_{Fin} + X_{Degree} + X_{SocInt} $&0.85 $\pm$ 0.002&0.56 $\pm$ 0.003&0.80 $\pm$ 0.002&0.47 $\pm$ 0.003\\

S03&$X_{Degree} + X_{SocInt} $&0.60 $\pm$ 0.002&0.16 $\pm$ 0.002&0.61 $\pm$ 0.003&0.18 $\pm$ 0.002\\
\hline
\end{tabular}}
\end{table}

The comparison of performance obtained by the models trained with synthetic data against the models trained on real-world data is presented in Table \ref{p03tbl:fake_real}. 
We can understand this comparison as the cost of using synthetic data, and it corresponds to the loss of predictive power to preserve the borrower's privacy. We can observe that in the best cases, this decrease in predictive power is approximately 3\% and 6\% when we measure the performance in AUC and KS, respectively.

\begin{table}[ht]
\footnotesize
\centering
\caption{Comparison between models trained using synthetic data and models trained on real data. $^{**}$ denotes when the difference is statistically significant using 0.05 as the p-value threshold, while $^{*}$ uses 0.1.}
\label{p03tbl:fake_real}
\resizebox{0.95\textwidth}{!}{%
\begin{tabular}{|c|c||c|c||c|c|}
\hline
\multirow{2}{*}{\parbox{2.0cm}{\centering Synthesizer Experiment}}&\multirow{2}{*}{\parbox{4.0cm}{\centering Classifier training features}}&\multicolumn{2}{c||}{ {\centering holdout 2018}}&\multicolumn{2}{c|}{ {\centering holdout 2019}}\\ \cline{3-6}
&&AUC diff&KS diff&AUC diff&KS diff\\ \hline

S01&$X_{Fin}$&-3.59\%$^{**}$&-10.09\%$^{**}$&-0.86\%$^{**}$&-3.92\%$^{**}$\\
S02&$X_{Fin}$&-6.24\%$^{**}$&-13.24\%$^{**}$&-3.32\%$^{**}$&-6.48\%$^{**}$\\
S03&$X_{Fin}$&-2.81\%$^{**}$&-6.01\%$^{**}$&-3.21\%$^{**}$&-6.70\%$^{**}$\\ \hline
S03&$X_{Fin} + X_{Degree} + X_{SocInt} $&-3.12\%$^{**}$&-5.68\%$^{**}$&-2.54\%$^{**}$&-4.73\%$^{**}$\\
S03&$X_{Degree} + X_{SocInt} $&-1.85\%$^{**}$&-4.31\%$^{**}$&-0.69\%$^{**}$&1.10\%$^{*}$\\ \hline

\end{tabular}}
\end{table}

\section{Conclusions}\label{p03sec:Conclusions}

This work aimed to use synthetic data to train creditworthiness assessment models. We used a massive dataset of 1 million individuals and trained state-of-the-art synthesizer methods to obtain synthetic data and achieve this goal. Then, we presented a training framework that allows us to analyze trained models with synthetic data and observe their performance on real-world data. In addition, we observed their performance one year after being trained to see how susceptible they are to data drift.
Our results show that lower quality synthetic data is obtained as we increase the number of attributes in the synthesizer. Despite this, it is possible to use these data to train models that obtain good results in real-world scenarios, with only a reduction in the predictive power of approximately 3\% and 6\% when we measure the performance in AUC and KS, respectively.
These findings are of great relevance since they allow us to train accurate creditworthiness models. At the same time, we keep borrowers' privacy and encourage financial institutions to strengthen ties with academia and foster collaboration and research in credit scoring without the privacy and security restrictions.

\section{Future Work}\label{p03sec:FutureWork}

Our future work will delve into how to synthesize social interactions' information in the form of graphs and not as added attributes to the training dataset since, as we show, this deteriorates the quality of the synthetic data.

\section*{Acknowledgements} This work would not have been accomplished without the financial support of CONICYT-PFCHA / DOCTORADO BECAS CHILE / 2019-21190345. The second author acknowledges the support of the Natural Sciences and Engineering Research Council of Canada (NSERC) [Discovery Grant RGPIN-2020-07114]. This research was undertaken, in part, thanks to funding from the Canada Research Chairs program.
The last author thanks the support of  MICIN UNDER project PID2020-116346GB-I00. 

%
%

\bibliographystyle{apalike}

\begin{thebibliography}{}

\bibitem[Akoglu, 2018]{Akoglu2018}
Akoglu, H. (2018).
\newblock User's guide to correlation coefficients.
\newblock {\em Turkish Journal of Emergency Medicine}, 18.

\bibitem[Bradley, 1997]{Bradley1997}
Bradley, A.~P. (1997).
\newblock The use of the area under the roc curve in the evaluation of machine
  learning algorithms.
\newblock {\em Pattern recognition}, 30(7):1145--1159.

\bibitem[Djeundje et~al., 2021]{Djeundjie2021}
Djeundje, V.~B., Crook, J., Calabrese, R., and Hamid, M. (2021).
\newblock Enhancing credit scoring with alternative data.
\newblock {\em Expert Systems with Applications}, 163:113766.

\bibitem[Fiore et~al., 2019]{Fiore2019}
Fiore, U., {De Santis}, A., Perla, F., Zanetti, P., and Palmieri, F. (2019).
\newblock Using generative adversarial networks for improving classification
  effectiveness in credit card fraud detection.
\newblock {\em Information Sciences}, 479:448--455.

\bibitem[Flach, 2012]{Flach2012}
Flach, P.~A. (2012).
\newblock {\em Machine Learning - The Art and Science of Algorithms that Make
  Sense of Data}.
\newblock Cambridge University Press.

\bibitem[Friedman, 2001]{Friedman2001}
Friedman, J.~H. (2001).
\newblock Greedy function approximation: a gradient boosting machine.
\newblock {\em Annals of statistics}, pages 1189--1232.

\bibitem[Gici{\'c} and Subasi, 2019]{Gicic2019}
Gici{\'c}, A. and Subasi, A. (2019).
\newblock Credit scoring for a microcredit data set using the synthetic
  minority oversampling technique and ensemble classifiers.
\newblock {\em Expert Systems}, 36(2):e12363.

\bibitem[Goh and Lee, 2019]{Goh2019}
Goh, R.~Y. and Lee, L.~S. (2019).
\newblock Credit scoring: a review on support vector machines and metaheuristic
  approaches.
\newblock {\em Advances in Operations Research}, 2019.

\bibitem[Goodfellow et~al., 2016]{Goodfellow2016}
Goodfellow, I., Bengio, Y., and Courville, A. (2016).
\newblock {\em Deep Learning}.
\newblock MIT Press.
\newblock \url{http://www.deeplearningbook.org}.

\bibitem[Hagberg et~al., 2008]{Hagberg2008}
Hagberg, A., Swart, P., and SChult, D. (2008).
\newblock Exploring network structure, dynamics, and function using networkx.
\newblock In {\em In Proceedings of the 7th Python in Science Conference
  (SciPy}, pages 11--15. Citeseer.

\bibitem[Hodges, 1958]{Hodges1958}
Hodges, J. (1958).
\newblock The significance probability of the smirnov two-sample test.
\newblock {\em Arkiv f{\"o}r Matematik}, 3(5):469--486.

\bibitem[Hripcsak and Rothschild, 2005]{Hripcsak2005}
Hripcsak, G. and Rothschild, A.~S. (2005).
\newblock {Agreement, the F-Measure, and Reliability in Information Retrieval}.
\newblock {\em Journal of the American Medical Informatics Association},
  12(3):296--298.

\bibitem[Kennedy et~al., 2013]{Kennedy2013}
Kennedy, K., {Mac Namee}, B., Delany, S., O’Sullivan, M., and Watson, N.
  (2013).
\newblock A window of opportunity: Assessing behavioural scoring.
\newblock {\em Expert Systems with Applications}, 40(4):1372--1380.

\bibitem[Kingma and Welling, 2013]{Kingma2014}
Kingma, D.~P. and Welling, M. (2013).
\newblock Auto-encoding variational bayes.
\newblock {\em arXiv preprint arXiv:1312.6114}.

\bibitem[Lei et~al., 2020]{Lei2020}
Lei, K., Xie, Y., Zhong, S., Dai, J., Yang, M., and Shen, Y. (2020).
\newblock Generative adversarial fusion network for class imbalance credit
  scoring.
\newblock {\em Neural Computing and Applications}, 32(12):8451--8462.

\bibitem[McHugh, 2013]{McHugh2013}
McHugh, M.~L. (2013).
\newblock The chi-square test of independence.
\newblock {\em Biochemia medica}, 23(2):143--149.

\bibitem[Mu{\~n}oz-Cancino et~al., 2021]{munoz2021}
Mu{\~n}oz-Cancino, R., Bravo, C., R{\'\i}os, S.~A., and Gra{\~n}a, M. (2021).
\newblock On the combination of graph data for assessing thin-file borrowers'
  creditworthiness.
\newblock {\em arXiv preprint arXiv:2111.13666}.

\bibitem[Mu{\~n}oz-Cancino et~al., 2022]{munoz2022a}
Mu{\~n}oz-Cancino, R., Bravo, C., R{\'\i}os, S.~A., and Gra{\~n}a, M. (2022).
\newblock On the dynamics of credit history and social interaction features,
  and their impact on creditworthiness assessment performance.

\bibitem[Ngwenduna and Mbuvha, 2021]{Ngwenduna2021}
Ngwenduna, K.~S. and Mbuvha, R. (2021).
\newblock Alleviating class imbalance in actuarial applications using
  generative adversarial networks.
\newblock {\em Risks}, 9(3).

\bibitem[{\'O}skarsdóttir et~al., 2019]{Oskarsdottir2019}
{\'O}skarsdóttir, M., Bravo, C., Sarraute, C., Vanthienen, J., and Baesens, B.
  (2019).
\newblock The value of big data for credit scoring: Enhancing financial
  inclusion using mobile phone data and social network analytics.
\newblock {\em Applied Soft Computing}, 74:26 -- 39.

\bibitem[Park Seong~Ho, 2004]{Park2004}
Park Seong~Ho, Goo Jin~Mo, J. C.-H. (2004).
\newblock Receiver operating characteristic (roc) curve: Practical review for
  radiologists.
\newblock {\em kjr}, 5(1):11--18.

\bibitem[{Patki} et~al., 2016]{SDV2016}
{Patki}, N., {Wedge}, R., and {Veeramachaneni}, K. (2016).
\newblock The synthetic data vault.
\newblock In {\em 2016 IEEE International Conference on Data Science and
  Advanced Analytics (DSAA)}, pages 399--410.

\bibitem[Simumba et~al., 2021]{Simumba2021}
Simumba, N., Okami, S., Kodaka, A., and Kohtake, N. (2021).
\newblock Spatiotemporal integration of mobile, satellite, and public
  geospatial data for enhanced credit scoring.
\newblock {\em Symmetry}, 13(4).

\bibitem[{The Basel Committee on Banking Supervision}, 2000]{Basel2000}
{The Basel Committee on Banking Supervision} (2000).
\newblock Principles for the management of credit risk.
\newblock {\em Basel Committee Publications}, 75.

\bibitem[Torres, 2018]{Torres2018}
Torres, D.~G. (2018).
\newblock {\em Generation of synthetic data with generative adversarial
  networks}.
\newblock PhD thesis, Ph. D. Thesis, Royal Institute of Technology, Stockholm,
  Sweden, 26 November.

\bibitem[Wan et~al., 2017]{Zhiqiang2017}
Wan, Z., Zhang, Y., and He, H. (2017).
\newblock Variational autoencoder based synthetic data generation for
  imbalanced learning.
\newblock In {\em 2017 IEEE Symposium Series on Computational Intelligence
  (SSCI)}, pages 1--7.

\bibitem[Xu et~al., 2020]{LeiXu2020Thesis}
Xu, L. et~al. (2020).
\newblock {\em Synthesizing tabular data using conditional GAN}.
\newblock PhD thesis, Massachusetts Institute of Technology.

\bibitem[Xu et~al., 2019]{LeiXu2019}
Xu, L., Skoularidou, M., Cuesta{-}Infante, A., and Veeramachaneni, K. (2019).
\newblock Modeling tabular data using conditional {GAN}.
\newblock {\em CoRR}, abs/1907.00503.

\end{thebibliography}

\end{document}